\documentclass[aps,prl,showpacs,superscriptaddress,twocolumn,floatfix,nofootinbib]{revtex4}

\usepackage{graphicx}
\usepackage{mathrsfs}
\usepackage{amsmath}
\usepackage{mathrsfs}

\newcommand{\be}{\begin{equation}}
\newcommand{\ee}{\end{equation}}
\newcommand{\ba}{\begin{eqnarray}}
\newcommand{\ea}{\end{eqnarray}}
\newcommand{\bs}{\begin{subequations}}
\newcommand{\es}{\end{subequations}}

\newcommand{\bac}{\begin{equation}
    \begin{array}{rcl}}
\newcommand{\eac}{\end{array}\end{equation}}

\newcommand{\forget}[1]{\iffalse#1\fi}
\newcommand{\forgetmenot}[1]{\iftrue#1\fi}

\renewcommand{\d}{\mathrm{d}}

\newcommand{\tot}{_{\mathrm{tot}}}
\newcommand{\eff}{_{\mathrm{eff}}}

\newcommand{\msl}{\mathscr{L}}
\newcommand{\msf}{\mathscr{F}}
\newcommand{\msg}{\mathscr{G}}

\begin{document}

\title{Photon gas dynamics in the early universe}

\author{Chris~Clarkson}
\email{chris.clarkson@uct.ac.za}
\affiliation{Cosmology and Gravity Group,
  Department of Mathematics and Applied Mathematics,
  University of Cape Town, Rondebosch 7701, Cape Town, South Africa}
\affiliation{Institute of Cosmology and Gravitation, Mercantile house,
University of Portsmouth, Portsmouth, PO1 2EG, United Kingdom}

\author{Mattias~Marklund}
\email{mattias.marklund@physics.umu.se}
\altaffiliation[Also at: ]{Centre for Fundamental Physics, Rutherford Appleton Laboratory, 
  Chilton, Didcot, Oxon OX11 OQX, U.K.}
\affiliation{Department of Physics, Ume{\aa} University, SE-901 87 Ume{\aa}, Sweden}

\begin{abstract}

Quantum electrodynamics predicts that photons undergo one-loop
scattering. The combined effect of this on the behaviour of a
photon gas for temperatures above $\sim10^{10}$~K results in a
softening of the equation of state. We calculate the effect this has on the effective
equation of state in the early universe, taking into account all
the species of the Standard Model. The change to the dynamics of the early
universe is discussed.


\end{abstract}
\pacs{}

\date{\today}
\maketitle

\paragraph*{Introduction}

The quantum vacuum has been a topic of fundamental interest in
physics since the advent of relativistic quantum theory, and has
spawned a number of experimentally verified concepts, of which a
few are the Lamb shift, Delbr\"uck scattering, photon splitting,
and pair creation \cite{Marklund-Shukla}. In particular, the
possibility of probing its structure using state-of-the-art
experimental facilities has recently been the in the focus of a
number of works (see e.g.\ \cite{quantum vacuum} and discussions
therein). It is well-known that the properties of the quantum
vacuum can have important consequences for black hole (BH) physics
\cite{Hawking}, introducing the concepts of BH thermodynamics
\cite{Thermo}.
The properties of the quantum vacuum also affects the propagation
of photons, introducing a background dependent phase velocity $v$
of pencils of light, such that $v \lesssim c$. This is a result of
so called elastic photon--photon scattering in which virtual
electron--positron pairs mediate the collisions between photons.
Most of the discussions in the literature regarding the effects of
background fields on the null-cone structure has been limited to
the weak field case. This is relevant when possible experimental
setups are discussed. However, the propagation of photons on a
very strong background field yield interesting new effects.

We shall consider the full one-loop quantum electrodynamic (QED)
correction [given by the lagrangian (\ref{eq:lagrangian2}) below]
represented in terms of Feynman diagrams as
\begin{equation}
\includegraphics[width=0.09\columnwidth]{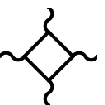}\,~ {}^{{}^{\displaystyle{+}}} \,
\includegraphics[width=0.09\columnwidth]{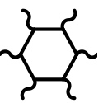}\,~ {}^{{}^{\displaystyle{+}}} \,
\includegraphics[width=0.09\columnwidth]{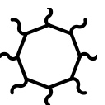}\,~ {}^{{}^{\displaystyle{+}}} \,
\includegraphics[width=0.09\columnwidth]{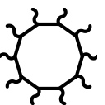}\,~ {}^{{}^{\displaystyle{+
\cdots}}}.\nonumber
\end{equation}
In Ref.\ \cite{Marklund-etal} it was shown that this contribution
 leads to a
nontrivial implicit dispersion relation for intense incoherent monochromatic 
photons. For super-critical intensities, i.e., for radiation
energy densities $\gtrsim
\epsilon_0|E_{\mathrm{crit}}|^2$, where $E_{\text{crit}%
}=m^{2}c^{3}/e\hbar \approx 1.32\times 10^{18}\,\mathrm{V/m}$
($\hbar $ is the Planck constant, $m_e$ is the electron mass, and
$c$ is the vacuum speed of light), is the critical field, this
leads to a noticeable decline in the photon phase velocity which for very
high energy densities ($\gg
\epsilon_0|E_{\mathrm{crit}}|^2$) approaches the value $v
\simeq c/\sqrt{5}$. Also, since the phase velocity is reduced, 
collective photon pair creation through the Schwinger mechanism 
is not possible even though the gas intensity exceeds the
corresponding Schwinger limit.
  It should be noted that for a very 
  hot thermal photon gas, a significant part of the Planckian distribution
  will be above the Compton frequency $m_ec^2/\hbar$, thus making single photon 
  pair production viable. 
In a cosmological context, the
result of such a nontrivial dispersion relation is to give an
effective equation of state (EOS) of radiation \cite{Marklund} of the
form $p_\gamma = \rho_\gamma/3n^2$, where $p$ is the isotropic
pressure, $\rho_\gamma$ is the radiation energy density, and $n =
c/v
\gtrsim 1 $ is the refractive index of the radiation (see also \cite{Partovi}).

In this paper we shall consider the weakening of the photon
equation of state on the evolution of the early universe. At the
onset of primordial nucleosynthesis, the effect of the change in
refractive index on the equation of state is small, but as
will be shown below such reductions in the effective radiation
pressure at earlier stages introduce noticeable effects.

\paragraph*{The photon gas equation of state}

The vacuum electromagnetic lagrangian for arbitrary field
strengths at the one-loop level is given by the Heisenberg--Euler
lagrangian~\cite{Schwinger}
\begin{eqnarray}
&&\!\!\!\!\!\!\!\!\!
 \msl = \tfrac{1}{2}\epsilon_0(a^2 - b^2) -\frac{\alpha}{2\pi}\epsilon_0E_{\text{crit}}^2\int_0^{\mathrm{i}\infty}
  \frac{dz}{z^3}\mathrm{e}^{-z}\times
\label{eq:lagrangian2} \\ &&\!\!\!\!\!\!\!\!\!
  \bigg[
  z^2\frac{ab}{E_{\text{crit}}^2}\,\coth\left(\frac{a}{E_{\text{crit}}}z\right)\,%
  \cot\left(\frac{b}{E_{\text{crit}}}z\right)
  - \frac{z^2}{3}\frac{(a^2 - b^2)}{E_{\text{crit}}^2} - 1 \bigg] ,
\nonumber
\end{eqnarray}
which is valid when the photon frequency satisfies $\omega \ll \omega_e \equiv m_ec^2/\hbar$.
Here,
$
  a = \left[(\msf^2 + \msg^2)^{1/2} + \msf\right]^{1/2} , \,
  b = \left[(\msf^2 + \msg^2)^{1/2} - \msf \right]^{1/2} ,
$ where  $\msf \equiv \tfrac{1}{4}F_{ab}F^{ab} =
  \tfrac{1}{2}(c^2\mathbf{B}^2 - \mathbf{E}^2) , \,
  \msg \equiv \tfrac{1}{8}\epsilon^{abcd}F_{ab}{F}_{cd} = -c\mathbf{E}\cdot\mathbf{B}$,
$\alpha$ is the fine-structure constant, $\epsilon_0$ is the
vacuum permittivity. The critical energy density is
$\rho_{\mathrm{crit}}=\epsilon_0|E_{\mathrm{crit}}|^2\approx1.55\times10^{25}\,$kg
m${}^{-1}$s${}^{-2}=(\pi^2k_B^4/15 \hbar^3c^3) T_\mathrm{crit}^4$
corresponding to a critical temperature $T_\mathrm{crit}\approx
1.2\times10^{10}\,
\mathrm{K}\approx 1.0\,$MeV. Hereafter, we shall use units where
$\hbar=k_B=c=1$.

As shown in~\cite{Marklund-etal}, the equation governing the
refractive index as a function of $\zeta=\rho_\gamma/\rho_{crit}$
is given by
\be\label{ri}
F(\sqrt{\zeta(n^2-1)}) =
\frac{4\pi}{\alpha}\frac{6\,(n^2-1)^2}{ n^2\,(5-n^2)},
\ee
where the function $F$ is
\be
F(x)
   =  \frac{1}{4\pi}
  \int_0^{\mathrm{i}\infty} \frac{dz}{z}\,\mathrm{e}^{-
  z/x} \left( \frac{1 - z\coth z}{\sinh^2 z} + \frac{1}{3}z\coth z
  \right) ,
\ee
where the integral is taken on a contour  to the right of the
imaginary axis. The solution to Eq.~(\ref{ri}) may be obtained by
considering the 
low energy approximation, $F(x)\simeq\frac{11}{45}x^2$, which will be sufficiently accurate for the effects considered here. As noted below Eq.\ (1), the governing Lagrangian is valid for arbitrary field strengths, but with the photon frequency $\omega \ll \omega_e$. For a thermal distribution, a significant number of the photons will have a frequency $\omega \gtrsim \omega_e$ when $T \sim T_{\text{crit}}$. Thus, the Planckian photon distribution implies that the present model i formally valid for $T < T_{\text{crit}}$. However, in what follows we will still make use of the Lagrangian (1) for temperatures $> T_{\text{crit}}$ in order to probe the consequences of such elastic photon--photon scattering. The high-$T$ model for this process, in a cosmological context, is a subject for future studies. 
We note that the refractive index satisfies $1< n<\sqrt{5}$, as may be seen
from Eq.~(\ref{ri}).

Using the low energy approximation, the equation of state parameter for a
photon gas, $w_\gamma(\zeta)=p_\gamma/\rho_\gamma=1/3n^2$, is the
solution of
\be
 \frac{w_\gamma\,(1-3w_\gamma)}{(15w_\gamma-1)} = \frac{11}{810}
 \left(\frac{\alpha}{4\pi}\right) \zeta \approx 7.9\times10^{-6}\zeta .\label{eos1}
\ee
Since $\zeta=(T/T_\mathrm{crit})^4$, we see from the small
dimensionless constant on the right hand side that photon-photon
scattering will start significantly affecting the dynamics of a
photon gas around $T\sim 10 T_\mathrm{crit}\sim 10\,$MeV. However, at such high temperatures, a black body spectrum will give rise to a significant pair production.
The rate of such pair and photon production can roughly be
modelled according to $\nu_{\gamma} = \sigma_{\gamma\gamma-ep}n_{\gamma}c$ and $\nu_{ep} = \sigma_{ep-\gamma\gamma}n_{e}c$, respectively. Here
$\sigma_{\gamma\gamma-ep}$ and $\sigma_{ep-\gamma\gamma}$ are the cross section for pair production and pair annihilation (in the center-of-mass frame), respectively, and $n_{\gamma}$ and $n_e$ are the photon and electron number densities, respectively. Such direct pair production by high frequency photons will immediately result in a nonlinear cascading process, dissipating the electromagnetic energy into fermionic degrees of freedom. It is possible to model this process more accurately using kinetic models, taking higher order interactions \cite{ruffini-etal}. However, a simple, but descriptive, local model is given by $\partial_tn_{\gamma} \approx \nu_{\gamma}n_{\gamma} - \nu_{ep}n_e$, $\partial_tn_e \approx -\nu_{\gamma}n_{\gamma} + \nu_{ep}n_e$, that gives rise to a typical initial exponential growth in the electron number density which reaches a nonlinear equilibrium at later times. Furthermore, the possibility of multi-photon pair production could be accounted for in a more accurate model.
However, the time-scale for pair-production for a high-temperature ($\gtrsim 1\,\mathrm{MeV}$) thermal photon distribution is short on a cosmological time-scale, 
and for our purposes it suffices to introduce a cut-off for the photon energy at $1-10\,\mathrm{MeV}$ (see Ref.\ \cite{ruffini-etal} for a discussion of the thermal range for pair production). Moreover, the onset of thermal equilibrium between photons and electron-positron pairs has only recently been investigated in detail using kinetic theory \cite{ruffini-etal}.
For
small temperatures, however, we have
\be
w_\gamma\approx\frac{1}{3}\left[1-\frac{22}{135}\left(\frac{\alpha}{4\pi}\right)\zeta\right],
\ee
which is actually a good approximation to $\sim5$MeV.

\paragraph*{Contribution to the overall EOS}

In the early universe, at temperatures above roughly one MeV, photons
are no longer a major contributor to the overall energy density.
As the temperature is increased above the rest mass of particles
in the Standard Model they become relativistic, contributing to
the energy density and pressure through the effective number of
degrees of freedom, $g_*(T)=(30/\pi^2)(\rho\tot/T^4)$, where
$\rho\tot$ is the total energy density, and $T$ is the temperature
of the photons~\cite{kt}.

The total energy density at temperature $T$ is given by a sum over
all species in equilibrium in terms of phase space
integrals~\cite{kt},
\be\label{rhotot}
\rho\tot=T^4 \sum_{i=\gamma,\nu,e,\mu,\cdots}
\left(\frac{T_i}{T}\right)^4
\frac{g_i}{2\pi^2}\int_{x_i}^{\infty}\d u \frac{(u^2-x_i^2)^{1/2}\, u^2}{\exp{u}\pm
1} ,
\ee
while the total pressure is given by
\ba
p\tot&=&w_\gamma\rho_\gamma \\
&+& T^4
\sum_{i=\nu,e,\mu,\cdots}
\left(\frac{T_i}{T}\right)^4
\frac{g_i}{6\pi^2}\int_{x_i}^{\infty}\d u \frac{(u^2-x_i^2)^{3/2}}{\exp{u}\pm
1} .\nonumber
\ea
Here, the $-$-sign refers to bosons, and the $+$-sign to fermions.
We have used $x_i=m_i/T$, and $g_i$ is the number of helicity
states of the $i$'th species. (We are assuming the chemical
potential is negligible.) Normally in the expression for the
pressure the photons may be included in the sum; here we have
written it separately in order to use the modified EOS,
Eq.~(\ref{eos1}). The energy density of the photons as a function
of temperature may be calculated from the appropriate term in
Eq.~(\ref{rhotot}), i.e., $\rho_\gamma=(\pi^2/15) T^4$, while the
modified pressure is given by the effective QED equation of state
$p_\gamma=w_\gamma\rho_\gamma$.

For some species the temperature $T_i$ may not be the same as the
photons, while still having a thermal distribution. Here, we
consider only the effects of neutrinos having this property: for
temperatures above a few MeV the neutrinos and photons share the
same temperature. When the electrons and positrons freeze-out
their entropy is transferred to the photons, raising their
temperature relative to the neutrinos by $T_\nu=(4/11)^{1/3}T$,
which may be found by conservation of
entropy~\cite{kt}.\footnote{It may be shown that the neutrino
freeze-out temperature is increased by around 0.1\% by the effects
considered here. }

We now define the effective equation of state parameter
\be
w\eff(T)=\frac{p\tot(T)}{\rho\tot(T)}
\ee
which is of course equal to $1/3$ for purely relativistic species.
However, as particles freeze out and become non-relativistic, the
contribution they make to $w\eff$ causes it to deviate from $1/3$
during the
freeze out. 
We shall analyse the effective equation of state as a means of
quantifying the QED correction to the photon gas equation of
state. We do this by first considering $w\eff$ in the standard
case of $w_\gamma=1/3$ and then compare it to $w_\gamma$ as given
by Eq.~(\ref{eos1}).

 The
difference between the equations of state is then given by
\be
w_\text{normal}-w_{QED}=\frac{2}{g_*(T)}\left(\frac{1}{3}-w_\gamma\right).
\ee
For low temperatures we find the correction to the standard EOS as
\be
w_\text{normal}-w_{QED}\simeq\frac{44}{405}\left(\frac{\alpha}{4\pi}\right)
\frac{1}{g_*(T)}\left(\frac{T}{1\,\mathrm{MeV}}\right)^{4}.
\ee
For temperatures around 1 MeV we have $g_*\approx 10.6$, and therefore find that the EOS is reduced by around 0.01\% at $T=1.5$MeV due to photon-photon scattering. This reaches 1\% by $T\simeq5$MeV, where we  must take into account the non-linear equation of state for $w_\gamma$ given by Eq.~(\ref{eos1});  by 10 MeV, when the
photons are contributing $\lesssim18\%$ of the total energy
density  we find that the EOS is reduced by more than 10\%. Of course, it is unlikely that the model we have used is sufficiently complex at such high temperatures. 

\paragraph*{Dynamics of the early universe}

Given our effective equation of state as a function of temperature
we may now calculate the dynamics of the early universe. Because
the density as a function of temperature is unchanged by
photon-photon scattering, the Hubble rate as a function of
temperature will be unchanged too, since
\be
H(T)^2=\left(\frac{1}{a}\frac{\d a}{\d t}\right)^2=\frac{8\pi
G}{3}
\rho\tot(T) .
\ee
The behaviour of the scale factor $a$, and time, $t$, with
temperature will be different between the two cases, since
$\dot\rho\tot=-3H\rho\tot(1+w\eff)$. Obviously this must be
calculated numerically in general. To demonstrate roughly how
photon-photon scattering modifies the evolution, let us define
$\varepsilon(T)=1/3-w\eff>0$. It may be shown that
$\varepsilon\lesssim0.05$ whatever EOS we use for the photons. To
first-order in $\varepsilon$ we find that the scale factor behaves
as
\be
\frac{a}{a_0}=\left(\frac{\rho\tot}{{\rho\tot}_0}\right)^{-1/4}
\exp\left[-\frac{3}{16}\int_{{\rho\tot}_0}^{\rho\tot}\frac{\d\rho\tot}{\rho\tot}\varepsilon\right].
\ee
The time evolution with temperature, on the other hand, is
\be
t=\frac{1}{2}\sqrt{\frac{3}{\kappa}}\rho\tot^{-1/2}-\frac{3}{16}\sqrt{\frac{3}{\kappa}}
\int_{{\rho\tot}_0}^{\rho\tot}\frac{\d\rho\tot}{\rho\tot^{3/2}}\varepsilon,
\ee
where $\kappa=8\pi G$. Therefore, we may estimate the  low
temperature change to the time and scale factor evolution to
leading order as
\ba
\frac{a_{QED}}{a_\text{normal}}&\sim&1-\frac{11}{540g_*}\left(\frac{\alpha}{4\pi}\right)\left(\frac{T}{T_\textrm{crit}}\right)^{4},\\
t_{QED}-t_{\text{normal}}&\sim&-\frac{11}{270 g_*^{3/2}}\sqrt{\frac{3}{\kappa}}
            \left(\frac{\alpha}{4\pi}\right)\left(\frac{T}{T_\textrm{crit}}\right)^{4}T^{-2}\nonumber,
\ea
where we have assumed that $g_*$ is constant. The evolution
of the scale factor evolves as normal at late times (low
temperature). Given the time difference between the two cases,  the time between, e.g.,
the quark-hadron phase transition and the onset of nucleosynthesis
will be shorter when photon-photon scattering is taken into
account.

\paragraph*{Discussion}

We have considered the collective effect that photon-photon
collisions (currently under experimental probing \cite{quantum vacuum}) 
have on the dynamics of the early universe, with
surprising results. Taking into account the full one-loop
Lagrangian for photon-photon scattering on an incoherent radiation
gas results in a drop in pressure of the photons. While the change in EOS is small at the onset of nucleosynthesis, the
change in the dynamics of the universe before this time could have
an effect on high-energy phenomena
such as primordial black hole production~\cite{xx}. However, more precise estimates
of such effects will require further research.

Photon-photon scattering should therefore be built into models
of the early universe in the hadron era and later. Indeed, as
primordial nucleosynthesis calculations are now accurate to
0.1\%~\cite{lt}, the fact that the neutrino temperature is
increased by this fraction imply that a further refinement may be
necessary in these models.

\acknowledgments We would like to thank Bruce Bassett for useful
discussions. Special thanks to Dominik Schwarz for crucial contributions.
CC is funded by the National Research Foundation (South Africa), and MM was funded by the Swedish Research Council.

\end{document}